\documentclass[10pt]{iopart}

\usepackage[linktocpage=true,colorlinks=true,pdfborder={0 0 0},linkcolor=blue,
citecolor=blue,filecolor=yellow,urlcolor=blue,bookmarks,pdfauthor={},]{hyperref}
\usepackage[T1]{fontenc}
\usepackage[latin9]{inputenc}
\usepackage[english]{babel}
\usepackage{xcolor}
\usepackage{graphicx}
\usepackage{dcolumn}
\usepackage{braket}
\usepackage{bm}
\usepackage[symbol]{footmisc}
\usepackage[nohyperlinks,nolist]{acronym}
\usepackage{booktabs}
\usepackage{multirow}
\usepackage{subfigure}

\newcommand{\ALCF}{Argonne Leadership Computing Facility, Argonne National Laboratory, Illinois 60439, USA}
\newcommand{\ICL}{Department of Materials, Imperial College London, London SW7 2AZ, UK}
\newcommand{\Basel}{Department of Physics, Universit\"at Basel, Klingelbergstr. 82, 4056 Basel, Switzerland} 
\newcommand{\Grenoble}{Univ.\ Grenoble Alpes, CEA, INAC-SP2M, L\_Sim, F-38000, Grenoble, France}

\begin{document}

\title{Accurate, Large-Scale and Affordable Hybrid-PBE0 Calculations with GPU-Accelerated Supercomputers} 

\author{Laura E.\ Ratcliff}     \address{\ALCF}\address{\ICL}\ead{laura.ratcliff08@imperial.ac.uk}
\author{A.\ Degomme}            \address{\Basel}
\author{Jos\'e A. Flores-Livas} \address{\Basel}
\author{Stefan Goedecker}       \address{\Basel}
\author{Luigi Genovese}         \address{\Grenoble}\ead{luigi.genovese@cea.fr}

\vspace{10pt}
\begin{indented}
\item[]December 2017
\end{indented}



\begin{abstract} 
Performing high accuracy hybrid functional calculations for condensed matter systems 
containing a large number of atoms is at present computationally very demanding 
-- when not out of reach -- if high quality basis sets are used.   
We present a highly efficient multiple GPU implementation of the exact exchange operator 
which allows hybrid functional density-functional theory calculations with systematic 
basis sets without additional approximations for up to a thousand atoms. 
This method is implemented in a portable real-space-based algorithm, released as an open-source package.
With such a framework hybrid DFT calculations of high quality become accessible on state-of-the-art 
supercomputers within a time-to-solution of the same order of magnitude as traditional semilocal-GGA functionals. 
\end{abstract}

\ioptwocol

\section{Introduction}

Density-functional theory (DFT) in principle is not an approximation, and should 
produce the exact ground-state energy and density, but in practice the crucial 
contribution, namely the exchange-correlation (XC) energy, is provided by an 
effective and therefore approximated, functional. Unsurprisingly, the quality of the 
result when compared to experiment depends on the quality of this approximation. 

A bibliographic search (see~\ref{app:biblio}) reveals that some 80 to 90 percent 
of all density-functional calculations conducted over the last 5 years in the chemistry 
community with Gaussian-basis codes such as {\sc Gaussian}~\cite{Gaussian} or 
{\sc NWChem}~\cite{NWchem} use highly accurate hybrid functionals. Among the most popular 
and widely used are B3LYP~\cite{B3,LYP}, PBE0~\cite{adamo1999-PBE0} and HSE06~\cite{HSE06} 
to name a few. These calculations use the best available functionals, however the total 
accuracy is limited by the size of the Gaussian basis set which cannot be taken to be very 
large in these calculations, because of the high computational cost. The most widely used 
basis set (in more than 90\% of the publications analyzed) in such hybrid functional 
calculations is the {\bf 6-31G} basis set which gives for instance atomization energy errors 
of about 30 kcal/mol~\footnote{See NIST website for the enthalpy of atomization (experimental values),  
http://cccbdb.nist.gov/ea1.asp}, i.e.\ about 30 times worse than the desired chemical 
accuracy~\cite{willand2013norm}, see for instance Jensen~{\it et al.}~\cite{jensen2017elephant}.  
It is not surprising that one of the most accurate basis sets, AUG-cc-pV5Z, which provides in most cases 
chemical accuracy, is used in less than 2 percent of the hybrid functional calculations because it is 
computationally prohibitive. 

In contrast to Gaussian-type basis sets, systematic basis sets such as plane waves or wavelets allow one to 
approximate the exact solution with arbitrarily high precision and with a moderate increase in cost within 
the adopted methodological framework. The framework is in this case the chosen exchange correlation functional 
together with the pseudopotential or PAW scheme~\cite{singh2006planewaves,RANGEL20161}. 
The latest generations of pseudopotentials and PAW schemes are able to deliver essentially 
chemical accuracy~\cite{willand2013norm} such that they do not compromise the overall accuracy of the calculation~\cite{Science}.  
Moreover the number of KS orbitals considered in all-electron quantum chemistry codes 
is much larger than in pseudopotential approaches, where one has to consider only valence and possibly semicore states.
By far the most popular codes based on systematic basis sets are the plane wave codes used 
preferentially in materials science and solid state physics such as 
{\sc Quantum-Espresso}~\cite{QE}, {\sc VASP},~\cite{vasp} and {\sc abinit}~\cite{gonze_abinit_2009}. 

However, less than 10 percent of the production results coming from these codes employ hybrid XC functionals. 
The main reason is that, for typical systems, hybrid functional calculations are {\it one to two orders 
of magnitude} more expensive than DFT-GGA calculations. It is therefore hardly possible, due to limited computer resources, 
to afford the computational power needed for such a treatment \emph{without introducing} additional approximations 
of the treatment, which might spoil the precision of the result in an unpredictable way. 
In other terms, even though accuracies that are very close to the desired chemical accuracy are 
in principle possible -- thanks to the availability of high quality basis sets as well as accurate exchange correlation functionals -- 
calculations that really attain this accuracy are still rare because of the high numerical cost for such simulations. 
Unsurprisingly the great majority of studies therefore still rely on local and semilocal LDA/GGA functionals. 

As anticipated, one way to circumvent the price that one has to pay for hybrid calculations is to exploit approximations 
or use a reduced representation of a set of KS orbitals. It has been shown by Gygi~\cite{gygi2009compact} that the algebraic 
decomposition of the matrix of wave-function coefficients provides a linear transformation which optimally localizes 
wave-functions on arbitrary parts of the basis set. This methodology drastically reduces the cost of hybrid-PBE0 calculations. 
However, the speed-up reached by deploying three-dimensional subspace bisection algorithms depends to a great extent 
on the degree of localization given by a tolerance $\epsilon$ that controls the quality of the results~\cite{gygi2012efficient}.    
More recently, Lin~\cite{lin2016adaptively} also used adaptive methods to compress the Fock-exchange operator, 
which also reduced the computational time associated with the calculation of this operator while not losing accuracy. 
However, these compressed methods are tailored for self-consistent field (SCF) optimizations based on a density-mixing scheme, 
on which the compression is performed for a given choice of the KS orbitals. This makes them unsuitable for use in 
less expensive SCF approaches like a direct minimization scheme, which are of utmost importance, for instance, 
in accurate and efficient {\it ab-initio} molecular dynamics trajectories. 
Thus, in order to study larger systems and more complex materials, we need to account for the requirements of a systematic basis 
set, computational affordability and algorithms without adjustable parameters.   
 
In this work, we present a highly efficient GPU implementation of a real-space based algorithm 
for the evaluation of the exact exchange, which reduces the cost of hybrid functional calculations 
in systematic basis sets, \emph{without any approximation}, by nearly one order of magnitude.

\section{Algorithm and Implementation}\label{lab:Theory}

We start this section with a brief introduction of the main quantities which will be the subject of analysis in the rest of the paper. 
In the (generalized) KS-DFT approach the one-body density matrix of the system is defined in terms of the occupied KS orbitals $\psi_i$:
\begin{equation}
F_\sigma({\bf r},{\bf r'}) = \sum_i f_{i,\sigma} \psi_{i,\sigma}^*({\bf r'}) \psi_{i,\sigma}({\bf r})\;,
\end{equation}
where we explicitly specify the (collinear) spin degrees of freedom with the index $\sigma=\uparrow,\downarrow$, 
together with the occupation number $f_{i,\sigma}$.
The system's electrostatic density is evidently the diagonal part of $F$, i.e.\ $\rho({\bf r}) = \sum_\sigma F_\sigma({\bf r},{\bf r})$.

The calculation of the exact exchange energy $E_X$ requires a double summation over all the $N$ occupied orbitals 
\begin{equation} \label{Eq:HF}
\eqalign{
E_X[\hat F]&=-\frac{1}{2} \sum_\sigma \int\mathrm{d}{\bf r} \: \mathrm{d}{\bf r}'
\frac{F_\sigma({\bf r},{\bf r'}) \: F_\sigma({\bf r'},{\bf r})} 
     {|{\bf r}-{\bf r}'|}\\
&=-\frac{1}{2} \sum_{i,j,\sigma} f_{i,\sigma} f_{j,\sigma} \int\mathrm{d}{\bf r} \: \mathrm{d}{\bf r}' \: 
   \frac{\rho_{ij}^\sigma({\bf r})  \: \rho_{ji}^\sigma({\bf r'}) } 
     {|{\bf r}-{\bf r}'|} \;,}
\end{equation}
where we have defined $\rho_{ij}^\sigma({\bf r}) = \psi_{j,\sigma}^*({\bf r}) \: \psi_{i,\sigma}({\bf r})$.
The diagonal ($i=j$) contribution to $E_X$ exactly cancels out the Hartree electrostatic energy $E_H[\rho]$.
The action of the Fock operator $\hat D_X$ to be added to the KS Hamiltonian directly stems from the $E_X$ definition:
\begin{equation}\label{daction}
\eqalign{
\hat D_X \ket{\psi_{i,\sigma}}&= \int\mathrm{d}{\bf r} \mathrm{d} {\bf r'} \:  \frac{\delta E_X[\hat F]}{\delta F_\sigma({\bf r},{\bf r'})} \psi_{i,\sigma}({\bf r'})\ket{{\bf r}} \\
&= -  \sum_{j} \int\mathrm{d}{\bf r}   f_{j,\sigma} V_{ij}^\sigma({\bf r}) \psi_{j,\sigma}({\bf r}) \ket{{\bf r}}    \;,}
\end{equation}
where we have defined 
\begin{equation}
V_{ij}^\sigma({\bf r}) = \int \mathrm{d} {\bf r'} \frac{\rho_{ji}^\sigma({\bf r'}) } 
     {|{\bf r}-{\bf r}'|} \;,
\end{equation}
that is the solution of the Poisson's equation $\nabla ^2 V_{ij}^\sigma = -4 \pi \rho_{ij}^\sigma$.
In a KS-DFT code which searches for the ground state orbitals, one has to repeatedly evaluate,
during the SCF procedure, for a given set of $\psi_{i,\sigma}(\mathbf r)$, the value of $E_X$ as well as the action of the corresponding Fock operator $\hat D_X$ on the entire set of occupied orbitals.

If each orbital $\psi_{j,\sigma}({\bf r})$ is written as a linear combination of $M$ Gaussian basis functions, 
a straightforward evaluation of the exact exchange energy has an $M^4$ scaling. 
Such a scaling might constitute a severe limitation for calculations with highly precise basis sets, 
where diffuse functions are needed to approach completeness. 
As a consequence the exact exchange becomes extremely expensive to calculate when very high quality atomic basis sets are used. 
Plane wave and wavelet basis set density-functional codes evaluate the exact exchange in a different way. 
They form all the $N (N+1)/2$ charge densities $\rho_{ij}^\sigma({\bf r})$ 
and then solve the Poisson's equation for each of them. Then the operator 
of Eq.~\ref{daction} can be evaluated as well as the value of 
\begin{equation}
E_X=-\frac{1}{2} \sum_{i,j,\sigma} f_{i,\sigma}f_{j,\sigma}
\int \mathrm d \mathbf r V_{ij}^\sigma(\mathbf r) \rho_{ji}^\sigma(\mathbf r) \;.
\end{equation}
The scaling of these operations is therefore $\mathcal O(N^2)$ multiplied by the scaling of the Poisson's equation, which is generally of $\mathcal O(N \mathrm{log} N )$ for typical Poisson solvers used in the community.

For large systems which exhibit the nearsightedness principle, exponentially~\cite{WCMS:WCMS1290} 
localized orbitals $\phi_\alpha(\mathbf r)$ can be constructed~\cite{nisanth} to represent 
the density matrix in terms of the matrix $\mathbf K$:
\begin{equation}
F(\mathbf r,\mathbf r') = \sum_{\alpha \beta} K^{\alpha \beta} \phi_\alpha(\mathbf r)
\phi_\beta(\mathbf r)\;,
\end{equation}
where we omit for simplicity the spin index and consider all the KS orbitals as real functions. 
After truncation within a localization region they can be used for the calculation 
of the Hartree exchange term.  In this case the matrix $\mathbf{K}$ becomes sparse so that the scaling reduces 
to a scaling proportional to $N$ with a very large prefactor:
\begin{equation}
E_X = \sum_{\alpha\beta\gamma\delta}
K^{\alpha \gamma} K^{\beta \delta}
\int \mathrm d \mathbf r V_{\alpha \beta}(\mathbf r) \rho_{\gamma \delta}(\mathbf r)
\end{equation}
More precisely the scaling is given by $N  \mathrm{log}(N) m^3$ where $m$ is the number of local orbitals for which the factor $K^{\alpha \gamma} K^{\beta \delta}$ is nonzero. 
The method we have developed could in principle also exploit orbitals constrained to localization regions, 
however we do not present such results since we are instead concentrating on optimization aspects for a given fixed simulation region. 

\subsection{Calculation Steps}

\begin{figure}[ht!]
\centering
\includegraphics[width=1.0\columnwidth]{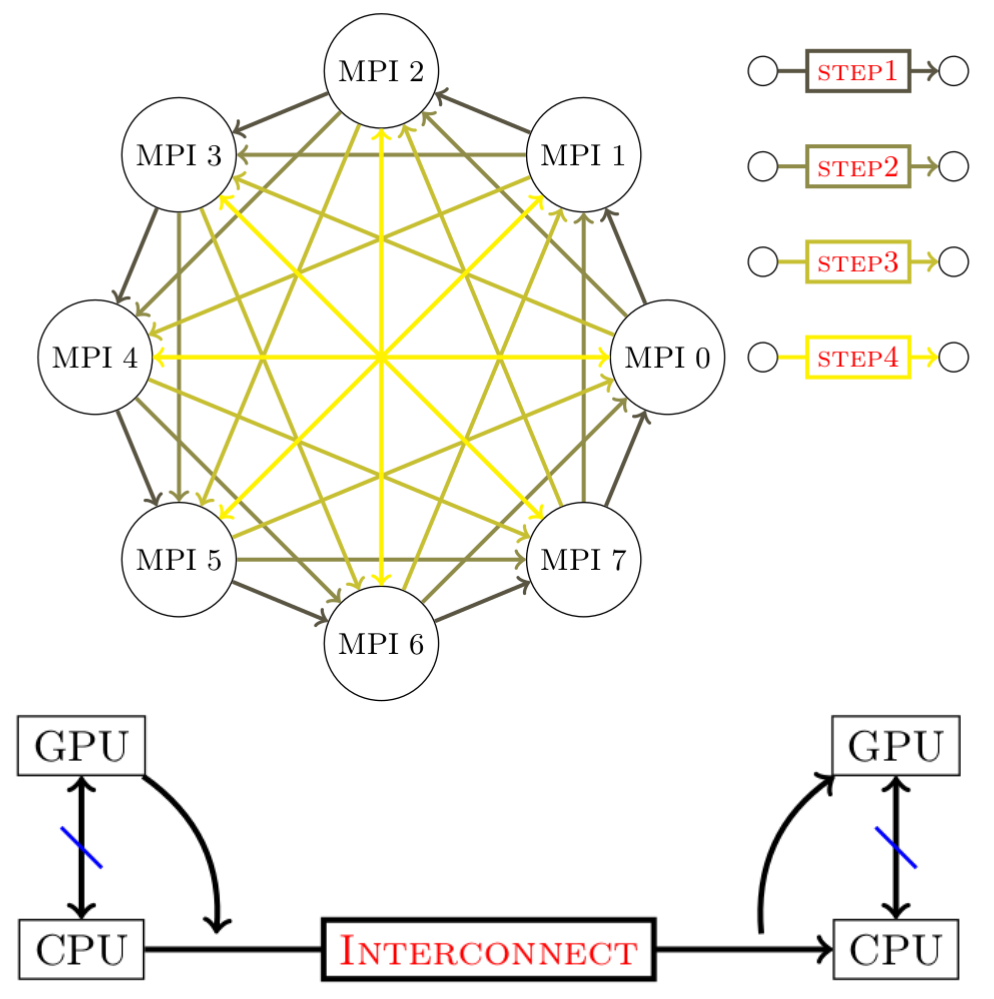}
\caption{~Sketch of the round-robin parallelization scheme used to calculate $E_X$ from Eq.~\ref{Eq:HF}. 
Each communication step is overlapped with the calculation of the Poisson's equation for the data 
that was communicated in the previous step. The advantage of the GPU-Direct scheme is illustrated in the bottom panel. 
GPU buffers are explicitly passed to MPI routines instead of sending data back and forth to the GPU.}
\label{fig:roundrobin}
\end{figure}

The starting point for evaluating the exact exchange terms (Eqs.~\ref{Eq:HF}, \ref{daction}) is the construction of the electrostatic potentials arising from the co-densities, 
\begin{equation}
\rho_{i,j}({\bf r}) = \psi_i^*({\bf r}) \: \psi_j({\bf r})\;,
\end{equation}
for each pair of orbitals $i$ and $j$. 
In the following we omit for simplicity the spin index $\sigma$.
We here describe the communication mechanism implemented to calculate the entire set of the $\rho_{i,j}({\bf r})$.

For a pool of $P$ MPI processes each labelled by $p=0,\cdots,P-1$  we assume that each process owns a subset of orbitals $\psi_{i_p}({\bf r})$.
Such a process is clearly able to calculate the $\rho_{i_p,j_p}$ codensities and the corresponding $V_{i_p,j_p}$. 
While performing such calculations, each process also sends (receives) the orbitals to (from) the $p\pm 1 \; \mathrm{mod}\: P$ process. Once this communication is terminated, the process $p$ is also capable of calculating $\rho_{i_p,j_{p-1}}$ and $V_{i_p,j_{p-1}}$. The latter transition potential is of interest at the same time for calculating the value of $E_X$ and the action of $\hat D_X$ on \emph{both} the orbitals $\psi_{i_p}$ and $\psi_{j_{p-1}}$. For this reason, once calculated, all the potentials $V_{i_p,j_{p-1}}$ are sent back to the $p-1$ process for the partial calculation of $\hat D_X \ket{\psi_{i_p-1}}$.
Such a procedure is repeated $[P/2]+1$ times, where at each step $s$ each process communicates with the $p\pm s \; \mathrm{mod}\: P$ MPI processes of the same pool. A schematic of this communication procedure is illustrated in Fig.~\ref{fig:roundrobin}. 

\subsubsection{The Calculation Kernels on Graphics Processing Units}
Since the orbitals represent substantial data packets the communication cost of this step is significant. 
Therefore, the communication is overlapped with the calculation using the data communicated at the 
previous step, thereby hiding the cost of the communication.  
The entire calculation procedure is decomposed into the following subprograms (kernels):
\begin{itemize}
\item Kernel 1: Calculate the charge density \\ $\rho_{i,j}({\bf r}) = \psi_i^*({\bf r}) \: \psi_j({\bf r})$ on a real space grid;
\item Kernel 2: Solve Poisson's equation $\nabla^2 V_{ij} = -4\pi \rho_{ij}$\;;
\item Kernel 3: Obtain the energy density and gradient by multiplying $V_{i,j}({\bf r})$ with $\rho_{i,j}({\bf r})$ or $\psi_j({\bf r})$ and $\psi_i({\bf r})$ respectively.
\end{itemize}
When a GPU accelerator is available, the above operations can be accelerated considerably with respect to the equivalent CPU kernels. 

For what concerns the communication scheduling, best performance was obtained using non-blocking 
Isend/Irecv MPI communications and by taking  advantage of the streaming capacities of GPUs, 
allowing the MPI library to perform the necessary tasks in the background. 
One-sided remote memory access (RMA) communications were evaluated and found to be slower on the tested machine and MPI implementations. 
This may however change depending on the communication libraries, the network or the tested cluster, 
and so this communication scheme can be turned on through an option in the input file if required. 
The Isend/Irecv scheme also has the advantage that it is applicable if using localized orbitals, which lead to unstructured communication patterns.

\subsubsection{Poisson Solver}
Most of the numerical work is done in kernel 2. 
For this step we use the established approach based on interpolating scaling 
functions~\cite{genovese2006efficient,Cerioni2013}, which has also been ported to GPU~\cite{dugan2013}.
The basic advantage of this approach is that the real-space values of the potential $V_{ij}$ 
are obtained with very high accuracy on the uniform mesh of the simulation domain, via a direct solution of 
Poisson's equation by convolving the density with the appropriate Green's function of the Laplacian. The Green's function can be discretized for the most common types of boundary 
conditions encountered in electronic structure calculations, namely free, wire, slab and periodic. 
This approach can therefore be straightforwardly used in all DFT codes that are able to express  
the densities $\rho_{ij}$ on uniform real-space grids. 
This is very common because the XC correlation potential is usually calculated on such grids.  
This approach has already turned out in its parallel CPU version to be fastest under 
most circumstances~\cite{garcia2014survey} and is therefore integrated in various DFT  
codes such as {\sc abinit}~\cite{gonze_abinit_2009}, {\sc CP2K}~\cite{hutter2014cp2k}, 
{\sc octopus}~\cite{andrade2012time} and {\sc Conquest}~\cite{CONQUEST_2007}. Recently this Poisson solver has been extended to also include electrostatic environments~\cite{Giuseppe1,Giuseppe2}.

The convolution is performed exactly with $N log(N)$ operations using Fourier techniques, in which appropriate padding is employed in order to avoid any supercell effects.
In our GPU implementation we use the cuFFT library from NVIDIA.  
This leads to an efficient implementation of the Poisson solver, as described in Ref.~\cite{dugan2013}.  
Such a GPU-based implementation of the Poisson solver already gives a gain in speed over the CPU
version, but the speed-up us limited by the large amount of host-GPU communication in such a simple
scheme where the data are stored on the host memory. 

\subsubsection{GPU Direct Communication}

Since large scale density-functional calculations are typically done on massively parallel computers, 
one clearly also has to exploit such architectures for hybrid functional calculations. 
Even if all the floating point operations are performed on a GPU, part of the efficiency would be lost 
in a parallel run if the data have to be transferred to the CPU for the MPI communication calls. 
For this reason we use the GPUDirect communication scheme which allows the MPI calls to be performed on the GPU. 

Host-GPU communications have always been one of the main issues with GPU acceleration. 
However, the Nvidia GPUDirect technology allows the acceleration of communication between GPUs by reducing or 
eliminating transfers between device memory, host memory and network buffers. 
It also provides an extension for MPI libraries, to use GPU memory directly in MPI communications, 
thus hiding the retrieving of the buffer from the user. 
The third version of GPUDirect, which is the most recent, allows the complete 
bypass of copies to the host memory, by connecting the GPU directly to the network card. 
This technology is currently available in most supercomputer centers, including the CSCS cluster {\sc Piz Daint}.  
Moreover, Cray's MPI implementation allows the use of this copy-free method without changing a CUDA-aware MPI code. 

Using this capability we have an almost perfect overlap of communication and computation, hiding data transfer involving the GPU. 
The GPU occupancy during the exact exchange step is over 80\%, with cuFFT kernels 
of the Poisson solver accounting for around 65\% of this GPU time. Using this optimized GPU 
to GPU MPI communication gives a speedup of three compared to the standard CPU based MPI 
communication used in the previous GPU implementation. 

\section{Tests and Case Studies}\label{lab:applications} 

\begin{figure}[h!]
\centering
\includegraphics[width=0.97\columnwidth,angle=0]{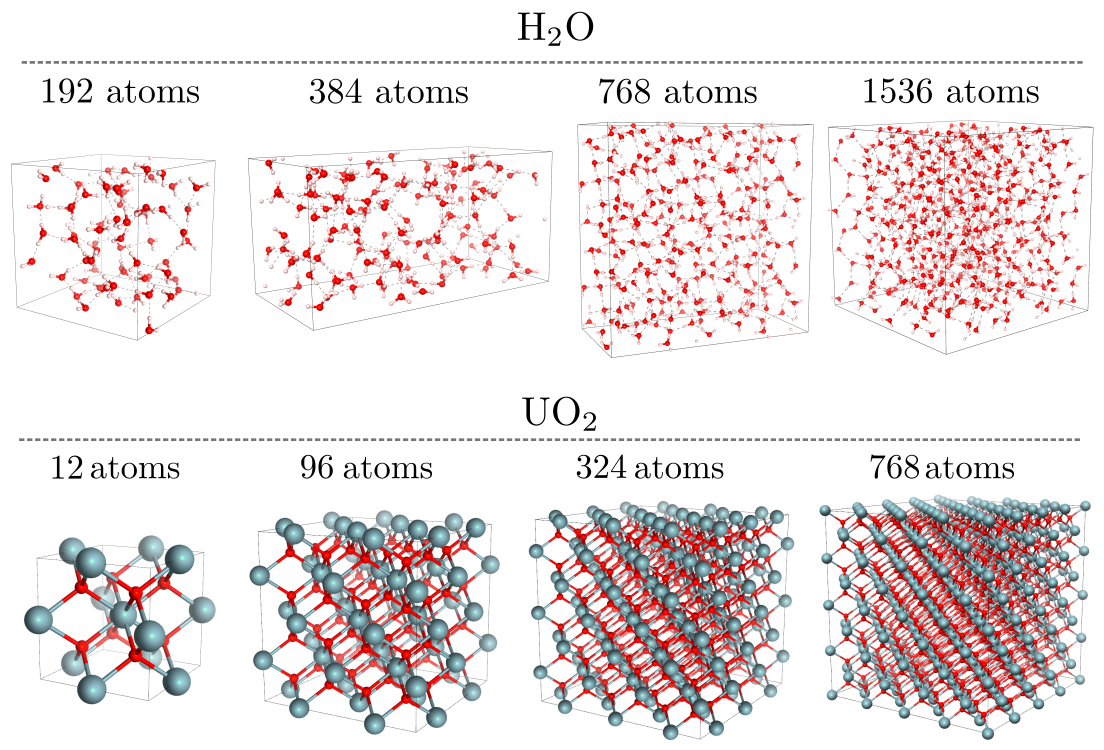}
\caption{~Systems used for benchmarking our implementation: water 
ranging from 64 to 512 molecules [top] and bulk UO$_2$ for system sizes of 12 atoms up to 768 atoms [bottom].}
\label{fig:Structures}
\end{figure}

We tested our approach on two different materials, to consider two opposing scenarios: 
a standard benchmark system with few electrons that allows us to put our results into context 
by comparing with other publications, and a highly challenging system with many electrons. 
For the former, we used H$_2$O, with sizes ranging from 64 to 512 molecules.  For the latter, 
we investigated uranium dioxide (UO$_2$), which plays a central role in nuclear fuel, and is 
both a technologically important and scientifically interesting material, with simulation challenges 
that go beyond the need for a hybrid functional, as discussed below. 
Fig.~\ref{fig:Structures} shows the structures used for UO$_2$ and H$_2$O.

All DFT calculations were performed with the {\sc BigDFT} 
code~\cite{genovese2006efficient,bigdft_2008,genovese2011daubechies}, which uses a systematic wavelet basis set. 
As the basis functions used in our calculation do not depend on the atomic positions of the systems, 
no Pulay terms have to be computed for the forces on the atoms. In other terms the calculation with hybrid functionals 
does not generate additional terms for the evaluation of the atomic forces other than the conventional Hellmann-Feynman terms.
The exchange-correlation functionals used were GGA-PBE~\cite{PBE96} and 
hybrid-PBE0~\cite{adamo1999-PBE0}. The pseudopotentials used were of the HGH form 
in the Krack variant~\cite{krack2005pseudopotentials}.  

In order to compare performance across different architectures, including both CPU 
and CPU-GPU, we employed three different parallel supercomputers. 
The machines used were the XC-30 and XC-50 {\sc Piz Daint} supercomputers 
hosted by the Swiss national supercomputing center (CSCS), Lugano, together with the 
{\sc Mira} Blue Gene/Q cluster at the Argonne Leadership Computing Facility (ALCF). 
For further details about the machines and calculation setups see~\ref{app:supercomp}.

To give an idea of the computational workload of our calculations we have written in Table~\ref{pscount} 
the number of evaluations of the Poisson solver that the calculation of $E_X$ and $\hat D_X$ 
(for a \emph{single} wavefunction iteration) requires for the systems considered in these benchmarks. 

To ease the understanding of our results, we use in the following the factor $\gamma$, 
defined as the ratio of walltime between a ground-state, self-consistent evaluation between a hybrid PBE0 
and a semilocal PBE ground state energy and forces evaluation. Since in our benchmarks the calculations 
are performed with the same number of wavefunction iterations such a factor is a reliable evaluation 
of the relative wall-time increase for PBE0 with respect to a PBE calculation. 
The reported timings in the following were obtained from {\sc BigDFT}'s internal profiling routines, 
which links to the standard library with real-time functions \texttt{librt}. 
These timings are in agreement with those obtained from Cray's Craypat tool 
and Nvidia profilers. 

\begin{table*}
\begin{tabular} {l c  c  c  c  | c  c c c c} \hline
  &\multicolumn{4}{c|}{H$_2$O}                     &           \multicolumn{5}{c}{UO$_2$}        \\ \hline \hline
 $N$& 192 & 384 & 768 & 1536 & 12 & 96 & 324 & 768 & 1029 \\
 \# $\psi_{i}^\sigma$ &256 & 512 & 1024 & 2048 & 164 & 1432 & 5400 & 12800 & 17150 \\ \hline
 \# $\rho_{ij}^\sigma$ & 32 896 & 131 328 & 524 800 & 2 098 176 & 6 658 & 513 372 & 7 292 700 & 81 926 400 & 73 539 200 \\ \hline
 \end{tabular}
 \caption{Number of Poisson solver evaluations per self-consistent iteration required to calculate the exact exchange energy and operator on the different systems used in the study. The number of atoms as well as the number of KS orbitals is indicated.} \label{pscount}
\end{table*}

\subsection{H$_2$O}

Internal coordinates of H$_2$O molecules were relaxed for different supercells and with both PBE and PBE0 functionals. 
Calculations were $\Gamma$-centered single $k$-point, using a grid spacing of $0.35$~\AA\ to reach convergence in the ionic forces of less than 2~meV/\AA. 
Pseudopotentials were constructed using the PBE approximation to the 
exchange and correlation with a single valence electron for hydrogen and 6 valence electrons for oxygen. 

\begin{table} 
\begin{tabular}{c c c c c | c}
H$_2$O    & Atoms  & Nodes  & PBE & PBE0 & $\gamma$\\  \hline
64        &  192   &    32  &  11 &   24 & 2.18     \\
128       &  384   &    64  &  23 &   74 & 3.2      \\
256       &  768   &   128  &  49 &  244 & 5.6      \\ 
512       & 1536   &   256  & 115 & 1030 & 8.95     \\
\end{tabular}
\caption{~PBE and PBE0 timings (s) obtained for complete SCF calculations of  H$_2$O systems as a function of the number of calculated molecules (first column) with Piz Daint (XC50).} \label{tab:agua}
\end{table}

\subsubsection{MD Energy Conservation}

\begin{figure}[h!]
\centering
\includegraphics[width=1.0\columnwidth,angle=0]{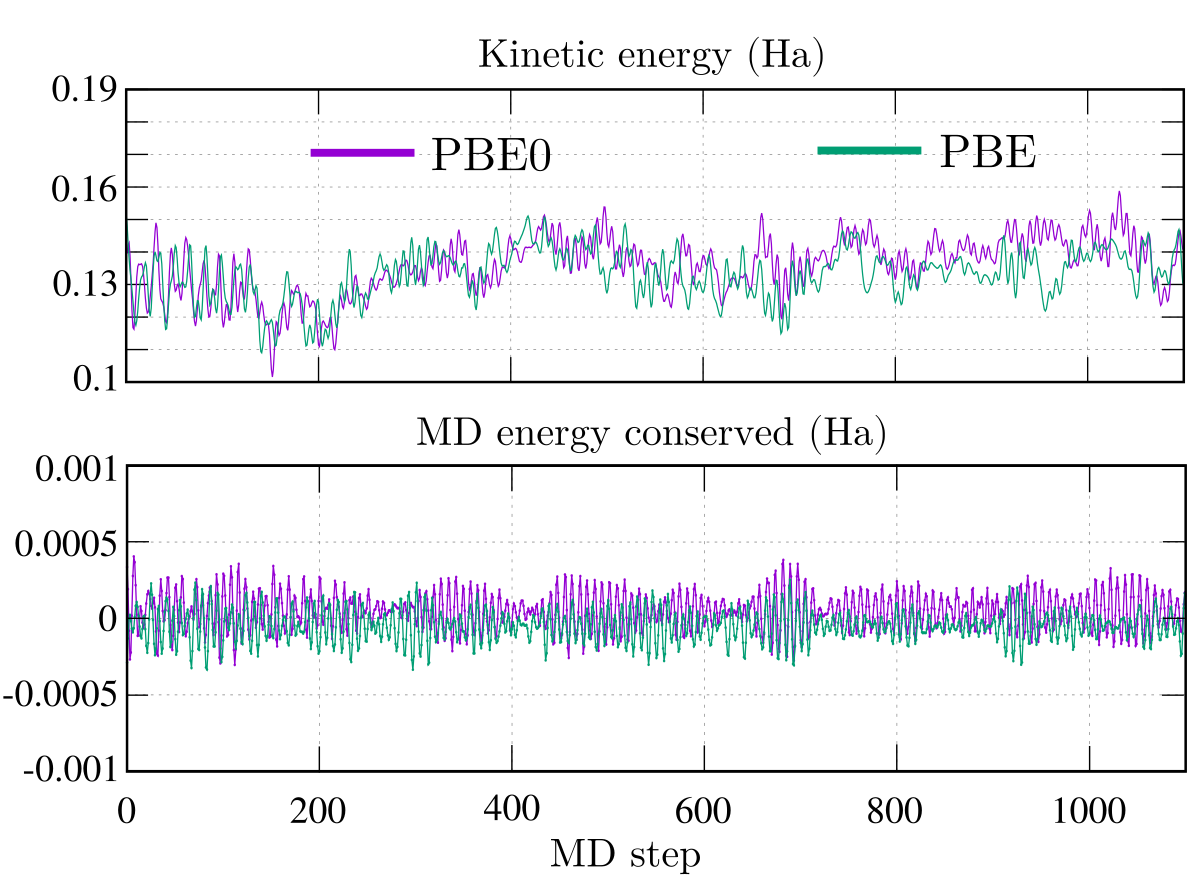}
\caption{~Kinetic energy [top] and normalised total energy [bottom] per MD step for PBE and PBE0 MD calculations of 64 H$_2$O molecules. Energy is conserved for a large number of steps for both PBE and hybrid PBE0.}
\label{fig:md}
\end{figure}

The first check was to verify the energy conservation in molecular dynamics (MD) trajectory calculations. 
This benchmark was conducted on 64 H$_2$O molecules (192 atoms) with a starting temperature of 
300~K and was evolved for more than 1000 steps on which wavefunction optimization was converged. 
These calculations represent a real production run with tight parameters and were conducted in Piz Daint (XC50). 
In Fig.~\ref{fig:md} the top panel shows the kinetic energy and the bottom panel shows the normalized total energy per MD-step. 
The energy is conserved for a large number of steps for both PBE and hybrid PBE0. The absence of energy drifting in the 
normalized energy is a clear evidence that energy and forces are accurately determined.

While the use of PBE does not represent a challenge for any {\it ab-initio} molecular dynamics code, 
the use of hybrid PBE0 has until now been impossible for such large systems that require too many MD steps. 
This is now feasible thanks to the great reduction in time for optimized MD steps.  
The PBE and PBE0 timings obtained for H$_2$O as a function of the number of calculated 
molecules are summarized in Table~\ref{tab:agua}. 
It is worth mentioning that the hybrid-PBE0 energies are accurately determined for a large number of atoms, while 
the absence of any adjustable parameters makes our approach suitable for tackling biological systems and nanostructures. 

\subsection{UO$_2$}

It is well known that standard Kohn-Sham DFT at the LDA or GGA level fails to handle self-interactions 
present in transition metal oxides, Mott Hubbard insulators and rare-earth materials. 
This situation is particularly bad for systems with partially occupied $d$ or $f$ shells, such as UO$_2$, and may lead to
incorrect (metallic) ground states for insulating systems. Indeed, the self interaction or delocalization error 
reduces the attractive potential, resulting in a rather extended orbital that increases its hybridization,  
therefore failing to capture the correct physics of the system.  
The problem is frequently circumvented by adding \emph{ad-hoc} terms to the KS Hamiltonian, 
modelling the on-site electronic iterations, whose strengths are usually called $U$ and $J$. 
However, these on-site interactions are adjustable parameters and it becomes more challenging 
to use this approach for predicting properties of materials. Hybrid functionals, on the other hand, which incorporate exact-exchange 
to the Kohn-Sham energy, also correct this self-interaction error without the need for additional, system-dependent parameters. 

Because of the relative computational costs, the vast majority of studies on UO$_2$ have thus far focused on DFT+U,
although there have been some hybrid calculations either for a full hybrid 
approach~\cite{Kudin2002,Prodan2006,Roy2008} or using the ``exact exchange for correlated electrons'' (EECE) approach~\cite{Novak2006,Jollet2009}. 
However these were restricted to small cells.

\subsubsection{Metastable states and Occupancy Control}

Aside from the need for a DFT+U or hybrid functional treatment for UO$_2$, another major complication arises, 
namely the existence of metastable states. These result from the many different combinations of orbitals that 
the two $f$-electrons of each uranium atom can occupy, which act as local minima.  
Both DFT+U and hybrid functional calculations easily become ``trapped'' in these local minima and therefore end 
up converging to different solutions for different initial guesses. 
These metastable states can have widely varying energies and band gaps, so that convergence to a metastable state 
can significantly affect the accuracy and reliability of the simulation.

A number of schemes for countering the problem of metastable states have been proposed, 
as summarized in Ref.~\cite{Dorado2013}.  One of the most popular is the occupation matrix control (OMC) scheme~\cite{Amadon2008,Jomard2008,dorado2009}, which is used to explore different metastable states 
by imposing various occupancy matrices. The systematic approach of the OMC forms the basis of its appeal, 
however it can require calculations on a rather large number of potential occupancy combinations, 
especially if not all uranium atoms are considered to be identically occupied or if non-zero off-diagonal elements 
of the occupation matrix are also considered. When combined with the use of large cells which are required when 
e.g.\ a defect is inserted, this provides a further strong motivation for the need to have a hybrid functional 
implementation with a low overhead.  

Previously, such an approach has been out of reach due to the excessive computational cost, 
however such calculations are possible using our new hybrid functional implementation. 
To this end, in a similar spirit to the occupancy control scheme of DFT+U, we have implemented an 
occupancy biasing scheme in \textsc{BigDFT}, which we use to ensure convergence to a particular state.

\begin{figure}[h!]
\centering
\includegraphics[width=0.97\columnwidth,angle=0]{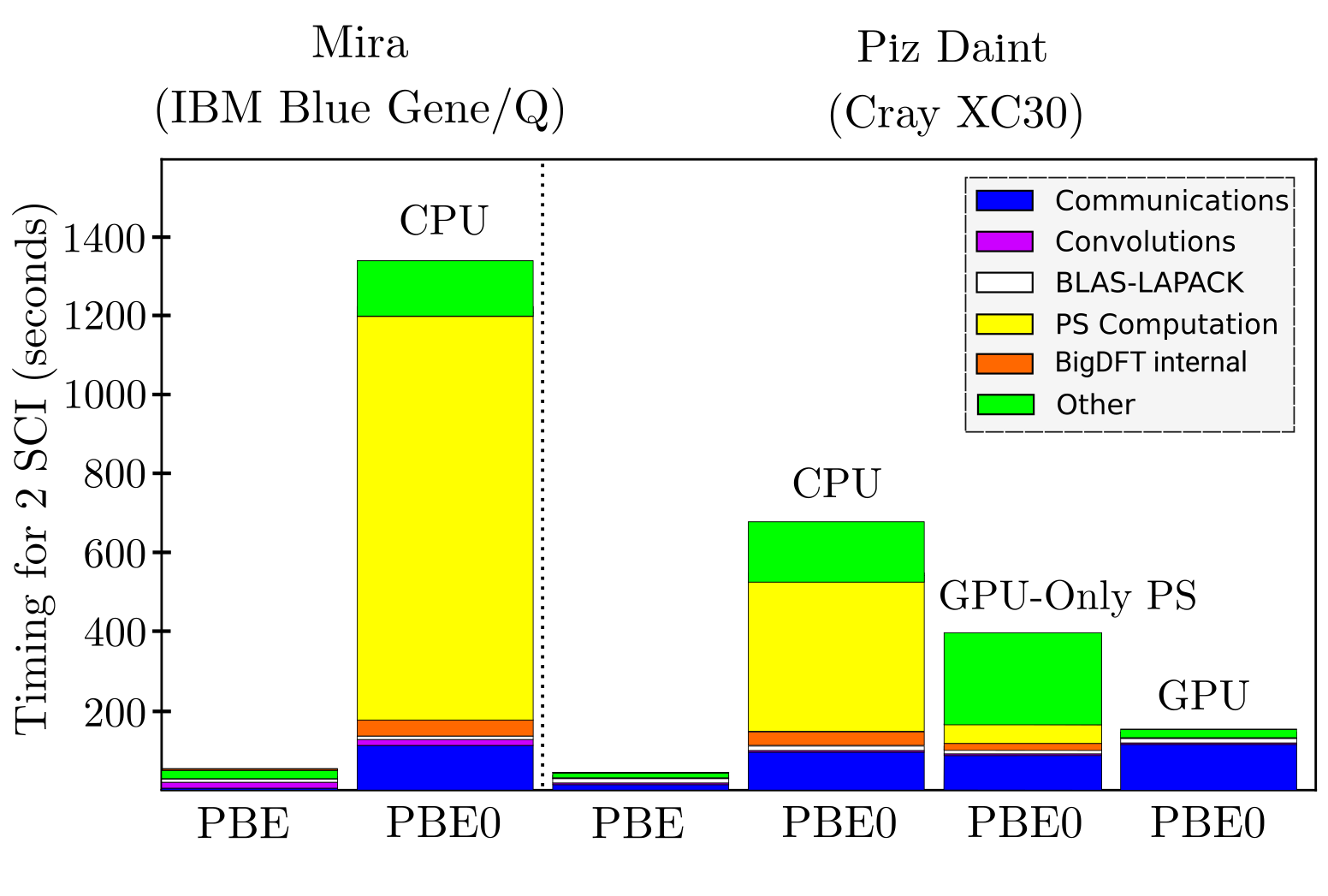}
\caption{~{\sc BigDFT} time profiling for 324 atoms of UO$_2$ (5,400 KS orbitals) on
a Blue Gene/Q ({\sc Mira}) and a Cray XC30 ({\sc Piz Daint}), both using 1,800 MPI processes.  
The calculations on Mira were performed using 4 MPI processes per node, with 16 OpenMP threads per MPI process. 
The solution of Poisson's equation in the exact exchange part consumes most of the time on the CPU architecture, 
whereas it is drastically reduced on the GPU accelerated architecture.}
\label{fig:cpu-gpu}
\end{figure}

\subsubsection{Computational Details}

We used a cubic cell, taking the structure from the Materials Project~\cite{MaterialsProject},
keeping the lattice constant fixed at 5.42~\AA.
In lieu of $k$-point sampling, we investigated increasing supercell sizes at the $\Gamma$-point only. 
We considered four different periodic cells covering a wide range of sizes:
a small cell (12 atoms), medium sized cell (96 atoms) and large cells (324 and 768 atoms).
Calculations employed a moderate grid spacing of $0.23$~\AA. 
The pseudopotential used for uranium had 14 electrons, while that for oxygen had 6 electrons. 

Restricting to the case of all atoms having the same $f$-orbital occupancy, we took the lowest energy configuration 
found by Krack~\cite{krack2015} subject to this constraint, namely $\left[f_{1}f_{3}\right]$. 
The occupancy was imposed for the first 12 self-consistent iterations (SCI), after which it was allowed to evolve freely.
Spin polarized calculations with non identical spin up and down orbitals were set for this material.
Following previous works, we used the 1-$\mathbf{k}$ anti-ferromagnetic ordering for the magnetic 
structure with no spin-orbit coupling, since the effects of spin-orbit coupling have been shown to be rather small~\cite{Roy2008}. 
No symmetry was imposed on the density.

A convergence threshold of $10^{-4}$ was used for the wavefunction gradient, 
except where only a fixed number of iterations were used for benchmarking purposes, as described when relevant. 
However calculations were not considered fully converged until the energy difference between two successive 
diagonalizations was less than 0.1\,meV/atom, with 12 SCI between each diagonalization.

\begin{figure*}[h!]
\centering
\subfigure[CPU-GPU: {\sc Piz Daint}]{\includegraphics[scale=0.55]{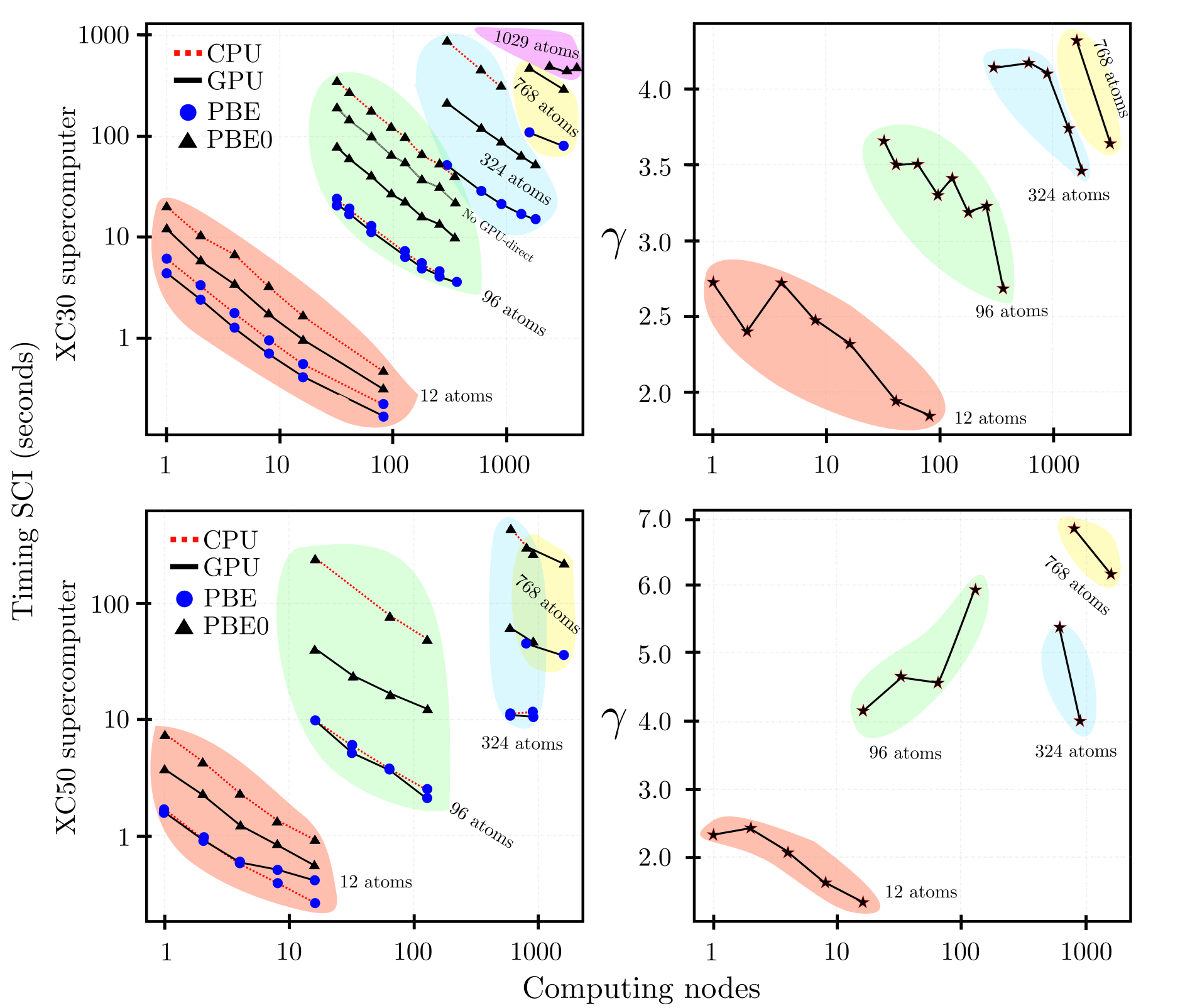}}
\subfigure[CPU-only: {\sc Mira}]{\includegraphics[scale=0.55]{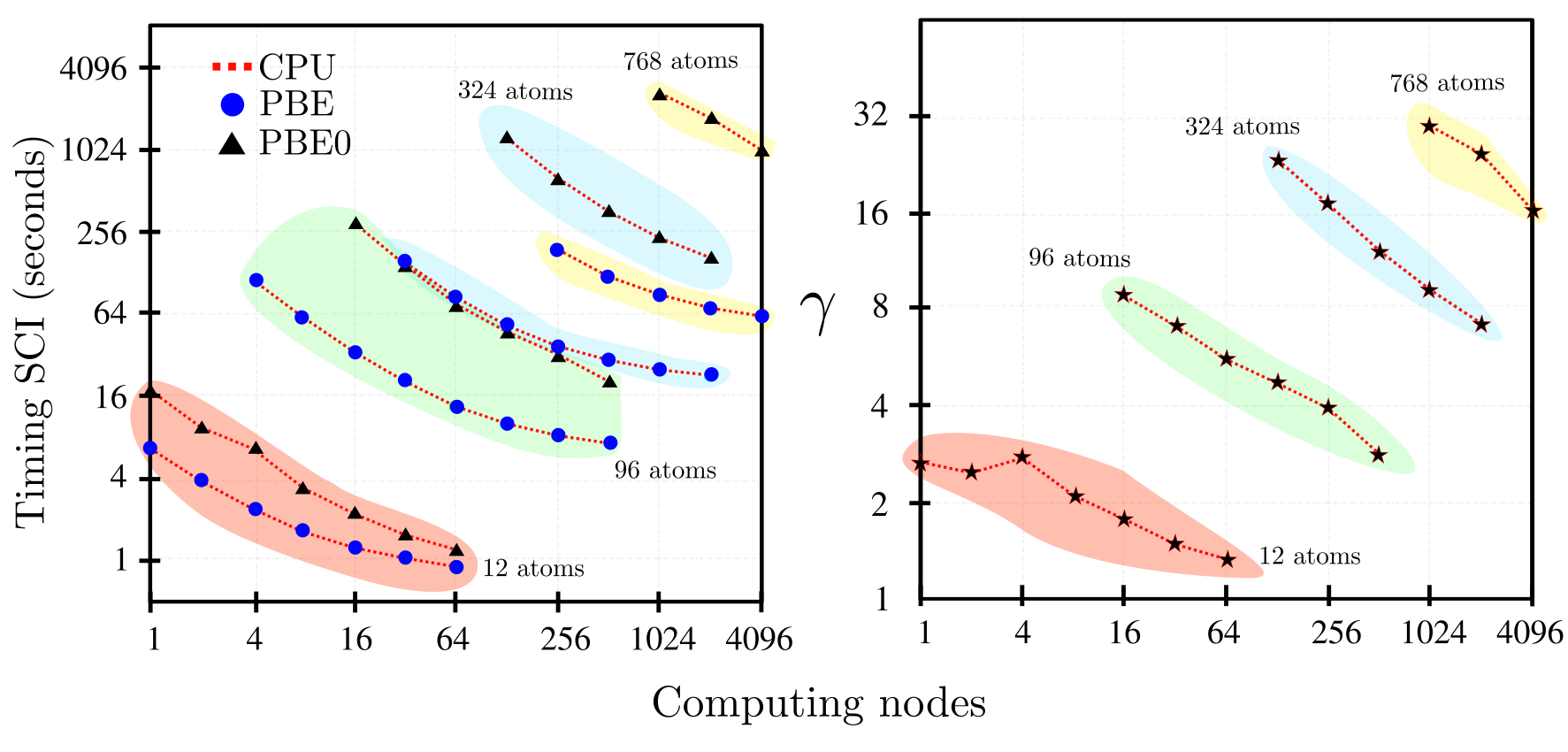}}
\caption{~Timings for PBE and PBE0 with and without GPU acceleration per iteration 
for different cells of UO$_2$ as a function of computing nodes on XC30 [top] and XC50 [middle]. 
For comparison, CPU-only results are also shown for {\sc Mira} [bottom].
Right panels represent the ratio PBE0/PBE for GPU runs [top and middle] and CPU runs [bottom].} 
\label{fig:performance}
\end{figure*}

\subsubsection{Performance and Scaling}

In order to analyze the performance of our approach, we first compared the breakdown in the timings for two SCIs 
on different architectures: CPU ({\sc Mira}) and CPU-GPU ({\sc Piz daint} XC30), as shown in Fig.~\ref{fig:cpu-gpu}. 
In this plot the $y-$axis represents wall-time in seconds and the color code categorizes the different computing 
operations necessary per iteration. This profiling was conducted for a system of 324 atoms of UO$_2$ (5,400 KS orbitals). 
The Nvidia profiler nvprof was used to profile the GPU kernels, and compute the efficiency of the implementation.  
We took timings for all the sections of {\sc BigDFT} that are related to the wavefunction optimization, not considering 
the time needed to set up the input wavefunctions and to evaluate the atomic forces as 
these operations do not involve calculations from our library. 
It is unsurprising that most of the time goes to the solution of Poisson's equation (PS computation) 
particularly on CPU architectures (yellow).

On XC30 systems exploiting only the CPU brings this down to 15 times the cost of PBE. 
Whilst the usage of GPU already brings the PBE0 computation within the same order of magnitude of the PBE runs, 
further reduction of the time required for the computation of PBE0 can be achieved by allowing GPU-Direct communication, 
in this case the cost to fully compute PBE0 on a systematic basis set for 324 atoms of UO$_2$ is only three times 
more expensive than for PBE.

We performed calculations for the different system sizes over a range of numbers of compute nodes.  
Fig.~\ref{fig:performance} shows the time per iteration for these calculations, for both the PBE and PBE0 functionals. 
For {\sc Piz Daint}, for 12 and 96 atoms, timings were obtained for runs where the wavefunction was fully converged. 
For the larger runs, as discussed above, we considered only two complete self-consistent iterations and the time is normalized per wavefunction iteration.  
For {\sc Mira} five SCI were considered for each system size. 
Each SCI still consists of a full calculation of both the exact exchange energy and gradient.
Table~\ref{tab:daint-mira} shows the ratios of PBE0/PBE for CPU and GPU architectures.  
The timings demonstrate not only the high scalability of our approach, but also show that in our GPU 
implementation the hybrid PBE0 calculation is just 2 times (for about 200 orbitals) and up to 4 times 
(for more than ten thousand orbitals) more time consuming than the GGA PBE calculation.  
In the CPU version we still have a relatively low ratio, i.e.\ a factor of 15 on {\sc Piz Daint}. 

\begin{center}
\begin{table*} \centering
\caption{~The ratio $\gamma$ obtained on {\sc Piz Daint} for CPU and GPU architectures, 
and on {\sc Mira} for CPU only. The number of orbitals considered varies due to a different number of unoccupied 
states used in the calculation.}
\begin{tabular} { c || l  l  c  c  c || l l c c} \hline
          &   \multicolumn{5}{c|}{\sc Piz Daint (XC30)}                     &           \multicolumn{4}{c}{\sc Mira}        \\ \hline \hline
 System   & Orbitals & Nodes   & Orbitals/ & $\gamma$ & $\gamma$&  Orbitals & Nodes   & Orbitals/ &  $\gamma$  \\
 (atoms)  &          &         &    Node   & CPU      &   GPU   &           &         & Process   &  CPU       \\   \hline
 {\bf 12} &    164   &     16  &  10       & 3.31     &   2.06  &   104     &    64	   &  1--2     & 1.33      \\
 {\bf 96} &   1,432   &   128  &  11        & 11.69    &   3.09  &   832    &   512   &  1--2     & 2.83     \\
{\bf 324} &   5,400   &   600  &  9        &  14.4    &   4.1   &   2,808    &  2,048  &  1--2     & 7.23      \\
{\bf 768} &  12,800  &  1,600  &  8        &   --     &   4.33  &   6,656   &  4,096 &  1--2        & 16.34   \\ \hline 
          \multicolumn{7}{c}{\sc Piz Daint (XC50)}  &  &  \\ \hline 
 {\bf 12} &    164    &  16     &  10       & 3.47   &   1.33  &   &   &   &   \\
 {\bf 96} &   1,432   &  128    &  11       & 19.76  &   5.91  &   &   &   &   \\
{\bf 324} &   5,400   &  600    &  9        &  40    &   5.41  &   &   &   &   \\
{\bf 768} &   12,800   &  1,600    &  8     &  -     &   6.14  &   &   &   &   \\ \hline 
\end{tabular}\label{tab:daint-mira}
\end{table*}
\end{center}

When going to two or even one orbital per node the degradation of the scalability is due to the fact that computation 
time is not high enough to overlap the communications. For one of the large systems used in this study 
(768 atoms, i.e.\ 12,800 orbitals),  the 3,200 node run (4 orbitals/node) was 75\% quicker than the 1,600 nodes run. 
This amounts to a 110~TFlops/s performance for the exact exchange computation. 
Even larger sizes (1,029 atoms, resulting in 17,150 orbitals, on up to 4,288 nodes) were reached during this study, 
although network issues arose and prevented a good scalability. 
These limits, although only reached on sizes that are currently considered as uncommon 
for scientific production work, could be overcome by adding a wavelet based compression/uncompression step before 
each communication.  This transformation is already implemented on the GPU in {\sc BigDFT}, 
and currently used only before entering the exact exchange step. On some systems this could 
reduce communication sizes by a factor of four.
The new version of {\sc Piz Daint} provides a huge bump in computing power on the GPU, 
while the network was only slightly improved. This means that this optimization is even more important.

During the exact exchange computation, when communication is properly overlapped, 
each of the GPUs spends 80\% of the time computing, and reaches 40~GFlops of 
sustained double precision performance. Memory throughput on the GPU is on average more than 165~GBps, 
over 65\% of the peak theoretical bandwidth of the GPU, which is the limiting factor here. 
The CPU performance for the same run without acceleration on {\sc Piz Daint} 
is 6.4~GFlops per CPU (using 8 OpenMP processes per CPU).

We have reported the performance results for the PBE0 hybrid functional. However quasi-identical timings for other 
hybrid functionals are expected to be of the same order of magnitude. 
We would also obtain identical timings for screened hybrid functionals such as 
HSE06, since in this case only the filter $F$ of kernel 2 would be different.  
In all cases the calculation of the Hartree exchange on the real-space grid would be done without any approximation.  

\subsubsection{Electronic Structure}

\begin{figure*}
\centering
\includegraphics[width=0.85\textwidth]{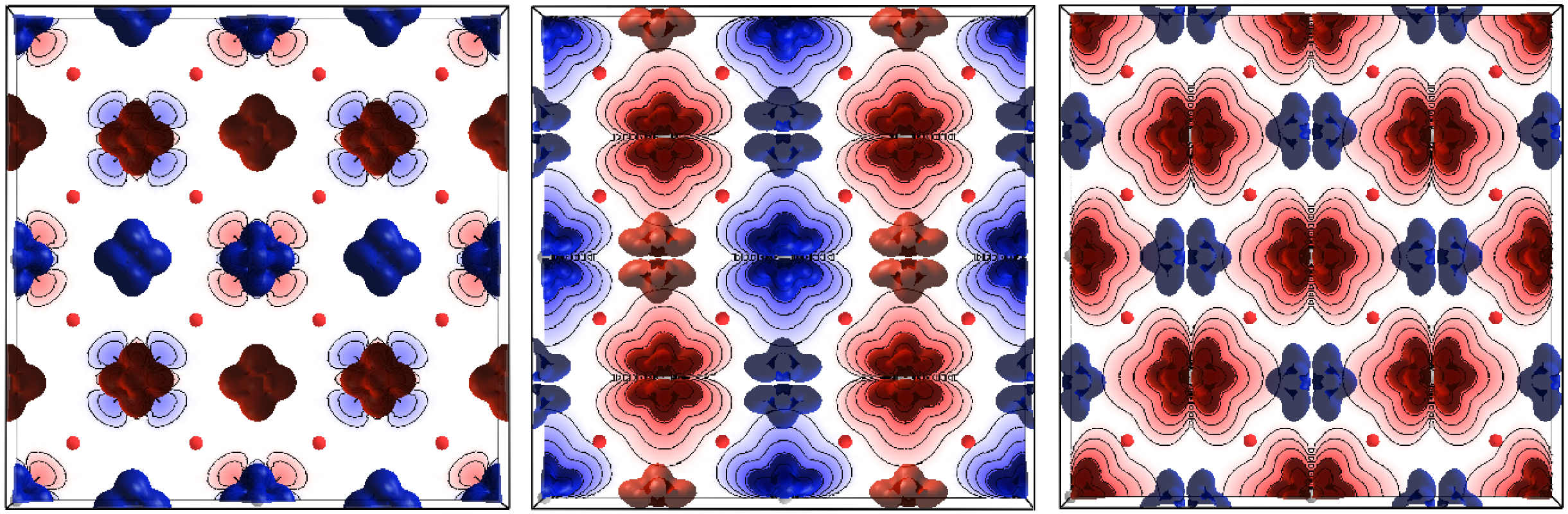}
\caption{~Final spin density for an occupancy of $\left[f_{1}f_{3}\right]$ in a $2 \times 2 \times 2$ supercell as viewed along the different axes. Oxygen (uranium) atoms are indicated in red (grey), while blue (red) density corresponds to spin up (down). 
The contour plots are shown for a single plane.}\label{fig:spindens}
\end{figure*}

To confirm that the occupancy biasing scheme is working as anticipated, we first printed the density matrix in the space of the $f$-orbital pseudopotential projectors at the end of the calculation, and verified that the correct states had non-zero values.
As a further visual confirmation, we plotted the spin density difference, as shown in Fig.~\ref{fig:spindens}, for the 96 atom cell.  
Aside from the expected rotation, the spin densities closely resemble those shown by Krack~\cite{krack2015} for the $\left[f_{-3}f_{-1}\right]$ 
state (which is degenerate with $\left[f_{1}f_{3}\right]$), 
confirming the ability of the occupancy biasing scheme to find the desired metastable state.

\begin{figure}
\centering
\includegraphics[width=0.49\textwidth]{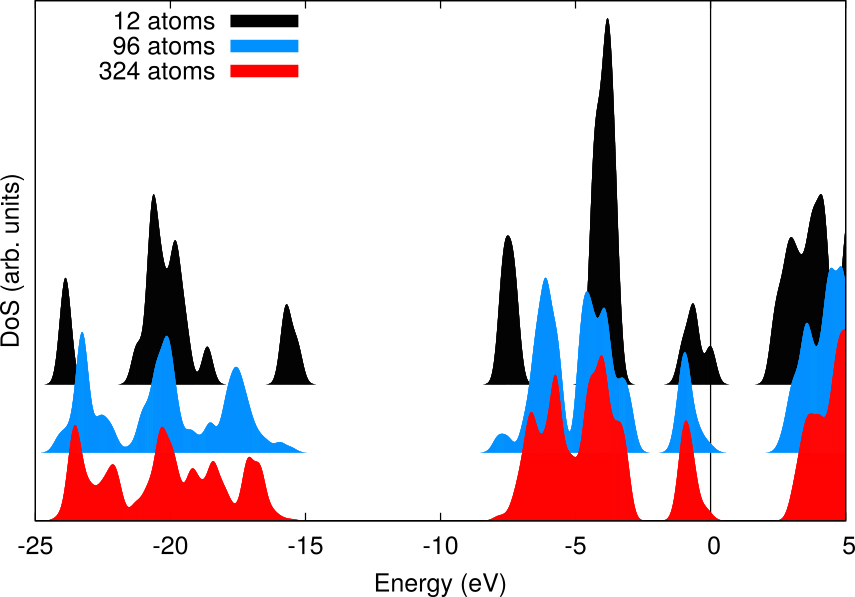}
\caption{~Density of states of UO$_2$ for spin up, for three different cell sizes, with an imposed occupancy of $\left[ f_{1}f_{3} \right]$. A Gaussian smearing of 0.2\,eV has been applied and the curves have been shifted so that the highest occupied molecular orbital (HOMO) energy of each curve is at zero. The curves have also been normalized with respect to the number of atoms in the cell. }\label{fig:dos_13}
\end{figure}

\begin{table}
\centering
\begin{tabular}{c c c c cc}
\toprule
Num. && \multirow{2}{*}{PBE0} && \multicolumn{2}{c}{DFT+U}  \\
atoms &&     && U$_{\mathrm{eff}}$=3.96 & U$_{\mathrm{eff}}$=2.00\\
 \cmidrule{1-1} \cmidrule{3-3}  \cmidrule{5-6} \\[-2.5ex]
12  && 2.35 && - & - \\
96  && 2.65 && - & - \\
324  && 3.01 && - & - \\
768  && - && 2.99 & 1.90 \\
  \bottomrule
\end{tabular}
\caption{~Band gaps in eV  for different supercell sizes of UO$_2$ for an imposed occupancy of $\left[ f_{1}f_{3} \right]$, calculated with PBE0.  For comparison, DFT+U results are also shown for two different values of U$_{\mathrm{eff}}$, using the values according to Krack~\cite{krack2015}. }
\label{tab:bandgap}
\end{table}

The density of states (DoS) is plotted for the different supercells in Fig.~\ref{fig:dos_13}, while the corresponding band gaps are given in Table~\ref{tab:bandgap}.  Only the DoS for up electrons is plotted, as results for spin down are indistinguishable.
Since no $k$-point sampling is used, it is unsurprising that the electronic structure and band gap are poorly converged for the smallest cell. 
The band gap is particularly affected by the supercell size, in line with previous observations where the level of $k$-point sampling was noted to affect the band gap accuracy~\cite{Jollet2009}.  Nonetheless, the differences in the DoS between the 96 atom and 324 atom supercells are much smaller.  Comparing the results with DFT+U, the calculated band gap for the largest supercell is very close to that according to Krack for an effective U value ($U_{\mathrm{eff}}=U-J$) of U$_{\mathrm{eff}}=3.96$, which is the value most commonly used in the literature. 

These results serve to demonstrate the ability of our approach to explore different $f$-electron occupancies in UO$_2$ using accurate hybrid functionals.  Furthermore, the computational cost is low enough to allow calculations on cell sizes which are big enough to treat point defects while keeping spurious interactions between defects in neighbouring cells to a minimum.  A detailed investigation of convergence with respect to basis size, the equilibrium lattice constant and exploration of the occupancy space using this approach will be published in the near future.

\section{Conclusions}\label{lab:conclusions}

Thanks to a highly efficient GPU implementation of a novel wavelet based algorithm for the evaluation of the exact exchange, 
we have demonstrated a reduction in the cost of hybrid functional calculations in systematic basis sets by nearly one order of magnitude. 
As a consequence, hybrid functional calculations with our method are only about three times more expensive than 
a GGA functional calculation. 
This is a price that the community is ready to pay for the significantly improved accuracy offered by such functionals. 

This methodology is available as a stand-alone library and can therefore be coupled to virtually any code 
that uses a systematic basis set and which calculates at some point the electronic orbitals on a Cartesian real space grid. 
Our implementation is completely general and available as an open-source package which can treat isolated and mixed 
boundary conditions as well as non-orthorhombic cells. 
As GPU acceleration is now also available on commodity computing clusters, our method will also allow 
researchers to do hybrid functional calculations for medium sized systems (hundreds of atoms) on affordable local computers. 
This advantageous price/performance aspect could also lead to an increased use of this technology in industry. 
Moreover, our developments will have implications both for future systems as well as for scientific applications. 
The availability of a code that gives a huge performance gain on accelerated architectures 
will further accelerate the spread of GPU accelerated systems. 
Finally, from a HPC viewpoint, the usage of such methods will 
enable extensive use of petaflop machines for electronic structure 
calculation communities, on the brink of the exascale era.

\section*{Acknowledgments}
This work was done within the MARVEL and PASC programs. 
We acknowledge computational resources from the Swiss National Supercomputing Center (CSCS) in Lugano (project s707 and 752). 
Computer time was also provided by the Innovative and Novel Computational Impact on Theory and Experiment (INCITE) program. 
This research used resources of the Argonne Leadership Computing Facility, which is a DOE Office of Science User Facility 
supported under Contract DE-AC02-06CH11357.

\appendix

\section{Supercomputer Details}\label{app:supercomp}

Performance was measured for runs conducted on three different parallel supercomputers. 
The system used to test our GPU accelerated runs is the {\sc Piz Daint} supercomputer 
hosted by the Swiss national supercomputing center (CSCS), Lugano. 
Benchmarks were performed on both the previous version of the cluster, and the upgraded version, deployed in December 2016. 
The original system machine was a 5,272 node Cray XC-30 system, which had 8-core Intel SandyBridge CPUs 
(Intel Xeon E5-2670), with 32~GB of RAM and an Nvidia Tesla K20X GPU with 6~GB of RAM per node. 
It used an Aries proprietary interconnect from Cray with a dragonfly network topology, and 
achieved a peak performance of 6,271~Teraflops. The compiler used for these runs was Intel 15.0.1.133, with CUDA 
toolkit 7.0. The MPI Library was Cray's MPICH 7.2.6. Each run was performed using 1 MPI process per node, 
and 8 OpenMP threads per MPI process.
The GPU portion of the new iteration of {\sc Piz Daint} is a 4,800 nodes Cray XC-50 system, with a 12-core 
(24 virtual cores with HyperThreading) Intel Haswell processor (Intel Xeon E5-2690 v3). 
Each node has 64~GB of RAM and is accelerated by an Nvidia Tesla P100 GPU with 16~GB of on board memory. 
This new system reached the peak performance of 9,779~Teraflops, due to the high increase in bandwidth at the GPU level. 
The network topology and interconnect is the same as on the former iteration of {\sc Piz Daint}. 
The compiler for these new runs was Intel 17.0.1, with CUDA toolkit 8.0. The MPI Library was Cray's MPICH 7.5.0.

The {\sc Mira} Blue Gene/Q cluster at the Argonne Leadership Computing Facility (ALCF) 
features 49,152 nodes and has a peak performance of 10~Petaflops. Each node consists of a 16 core 
PowerPC A2 1600 MHz processor, and 16GB of DDR3 memory. 
The network is an IBM 5D Torus Proprietary Network. The compiler used was IBM's xlc, version 14.1, 
with the provided MPI library, based on MPICH itself. 
Except where stated otherwise, runs were performed with 1 MPI process and 32 OpenMP threads per node, in order to maximize the range of node counts over which scaling tests could be performed.

\section{Bibliographic Search of DFT Calculations}\label{app:biblio}

The bibliographic search was conducted using the {\sc Google Scholar} engine in 2017. 
The procedure was as follows: the main papers suggested for citation in the 
{\sc Gaussian}~\cite{Gaussian} 2009 and {\sc NWChem}~\cite{NWchem} 2010 papers were used. 
Key words matching either in the abstract, title or block of the paper were accounted for. 
These results are shown in Table~\ref{tab:google}. About 80 to 90 percent of all density-functional calculations 
conducted with those two codes employed hybrid functionals. Not surprisingly these calculations 
use the best available functionals. 
The most widely used basis set in such hybrid functional calculations is the {\bf 6-31G} basis set which gives for instance 
atomization energy errors of about 30\,kcal/mol. 
The more accurate AUG-cc-pV5Z basis set, which give in most cases chemical accuracy, 
is used in less than 2 percent of the hybrid functional calculations. 
Clearly, while one problem is alleviated by using hybrid functionals, another problem is raised by employing poor basis sets.

\begin{center}
\begin{table*}[htb] \centering
\caption{~Approximate number of entries counted with the {\sc Google Scholar} engine for two widely 
used DFT codes in the chemistry community.}
\begin{tabular} { l  c | l  c} \hline
  \multicolumn{2}{c|}{{\sc Gaussian 09} Rev. A.02 (2009)}  & \multicolumn{2}{c}{{\sc NWchem} (paper 2010)}  \\ \hline
  \multicolumn{2}{c|}{Total calls: {\bf 22,835}}    & \multicolumn{2}{c}{Total calls: {\bf 1,455}}  \\  \hline
Keywords             & Citations  &  Keywords         & Citations    \\  \hline
"B3LYP" "6-31G"      & 11,000  &  "B3LYP"             & 882 \\
"B3LYP" "6-311G"     &  3,840  &  "Hybrid functional" & 274 \\
"B3LYP" "AUG-cc-pVDZ"&  1,950  &  "PBE0"              & {\bf 270} \\
"B3LYP" "AUG-cc-pV5Z"&    269  &  "HSE06"             & 19  \\ \hline
"PBE" "6-31G"        &  1,970  &  &  \\
"PBE" "6-311G"       &   855   &  &  \\
"PBE" "AUG-cc-pVDZ"  &   552   &  &  \\
"PBE" "AUG-cc-pV5Z"  &    80   &  &  \\ \hline
"PBE0" "6-31G"       & 1,540   &  &  \\
"PBE0" "6-311G":     &   607   &  &  \\
"PBE0" "AUG-cc-pvdz" &   425   &  &  \\
"PBE0" "AUG-cc-pV5Z" & {\bf 79}&  \\ \hline
  \end{tabular}\label{tab:google}
\end{table*}
\end{center}

\section*{References}

\bibliographystyle{iopart-num}
\bibliography{main}
\end{document}